\documentclass{article}
\usepackage[fontsize=9.8pt]{fontsize}
\usepackage{spconf,amsmath,amssymb,graphicx,url,times,color,booktabs,multirow,hyperref}
\usepackage[super]{nth}

\title{Rigid-Body Sound Synthesis with Differentiable Modal Resonators}
\name{Rodrigo Diaz, Ben Hayes, Charalampos Saitis, Gy{\"o}rgy Fazekas, Mark Sandler \thanks{Rodrigo Diaz and Ben Hayes are supported by UK Research and Innovation [grant number
EP/S022694/1].}}
\address{Centre for Digital Music, Queen Mary University of London}
\begin{document}
%
\maketitle
\begin{abstract}
Physical models of rigid bodies are used for sound synthesis in applications from virtual environments to music production.
Traditional methods such as modal synthesis often rely on computationally expensive numerical solvers, while recent deep learning approaches are limited by post-processing of their results.
In this work we present a novel end-to-end framework for training a deep neural network to generate modal resonators for a given 2D shape and material, using a bank of differentiable IIR filters.
We demonstrate our method on a dataset of synthetic objects, but train our model using an audio-domain objective, paving the way for physically-informed synthesisers to be learned directly from recordings of real-world objects.
\end{abstract}
\section{Introduction}
\label{sec:introduction}

\begin{figure*}[t]
    \centering
    \includegraphics[width=\textwidth]{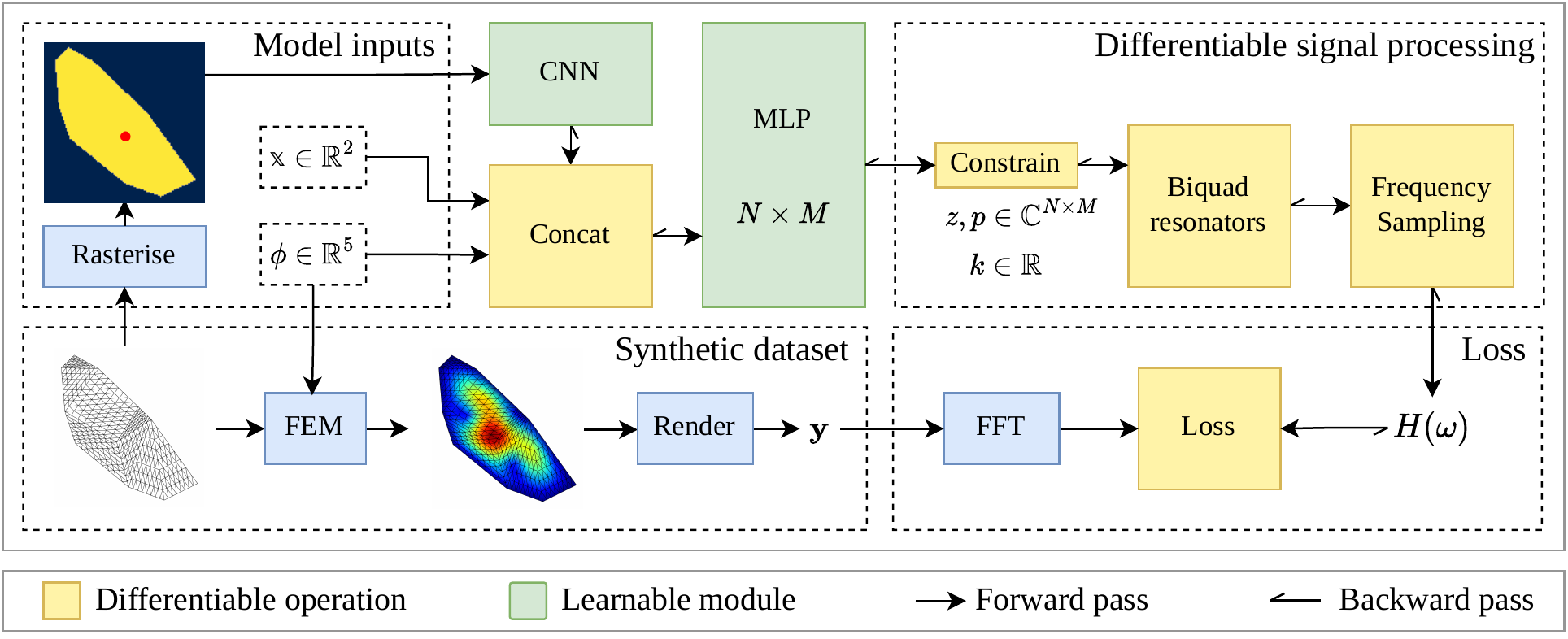}
    \caption{A schematic diagram of our model and its training pipeline. A CNN generates shape embeddings from a synthetic dataset of 2D convex shapes, which are concatenated with material parameters and excitation co-ordinates and passed to an MLP to generate parameters for a differentiable IIR filterbank.}
    \label{fig:pipeline}
\end{figure*}

The synthesis of contact sounds from rigid bodies and materials has been of continual interest in applications from music production, through sound design, to the rendering of object sounds in virtual environments \cite{o2002synthesizing, van2001foleyautomatic}.
For this reason, the problem has been extensively studied, and physically-based numerical methods are often selected for its solution.
However, such approaches typically incur significant computational cost, as well as storage cost for cached solutions if real-time interaction is required.
These limitations inhibit flexibility in such applications, where adapting to new object shapes and materials requires prohibitively time-consuming computation of the new solution.

Of these numerical approaches, which ultimately rely on solving the wave equation, the finite difference time domain and the finite element method (FEM) are most commonly used due to their adaptability.
FEM solvers are used to precompute the vibrational modes of arbitrarily shaped rigid objects, allowing contact sounds for these objects to be synthesized using modal synthesis. 

The computation of the modes is usually posed as a generalized eigenvalue problem using the mass and stiffness matrices of an object. 
Using the solution, sound can be rendered with an oscillator bank by projecting an impulse onto the modes at each discrete location within the object.
This approach to computing an object's modes can, however, be computationally expensive.
Moreover, every time the shape or material characteristics of the object change, the system must be solved again. 
Thus, numerous approaches have been proposed to accelerate this process~\cite{rausch2017modal,bonneel2008fast,chadwick2012faster}.

\subsection{Data-driven methods}

A recent generative method, proposed by Traer et al.~\cite{traer2019perceptually}, used a perceptually derived statistical model to approximate object impulse responses by approximating the modes within some degree of perceptual tolerance. While this method circumvents the need for a numerical solver and uses filters to account for different object interactions, it relies on abstract intermediate representations of an object's physical characteristics, and does not take into account object geometry.

Jin et al.~\cite{jin2020deep, jin2021deepeigen, jin2022neuralsound} trained a deep neural network to produce an object's modes, circumventing the need for a FEM solver. Their model predicted eigenvalues and eigenvectors from sparsely voxelized objects, using as supervision the corresponding modes obtained from the solver. Additionally, in order to generalize to different sizes and materials, the authors suggest the adaptive scaling of the modes based on material parameters (except for Poisson's ratio) and the size of the object. Similarly, our method uses a discretized shape representation as input to a neural network.
However, unlike Jin et al.'s model, no post-processing is required to account for different material parameters. Further, our model is not limited to querying modes at discrete positions in the shape, allowing for arbitrary co-ordinate input.




\subsection{Differentiable resonators}

While it is possible to synthesize modes using an oscillator bank where each damped sinusoid corresponds to an impulsed mode, resonant filter banks are commonly used for synthesis due to their flexibility, as they can be excited with a variety of signals~\cite{conan2014synthesis}.
Realising such a filter bank, however, requires the corresponding eigenvalue and eigenvector problems to be solved in order to derive the filter coefficients.
In this work, we propose an end-to-end learning method for generating resonant filter banks without explicitly solving the system.

Our method relies on the use of differentiable infinite impulse response (IIR) filters, which allow a neural network to directly control filter coefficients by propagating loss gradients through them. Whilst it is possible to directly specify such a filter in its recurrent form~\cite{kuznetsov2020differentiable, pepe2020designing}, this requires backpropagation of loss gradients through time which can introduce instability in the form of exploding or vanishing gradients, and is poorly suited to parallel computation on a GPU. 
For these reasons, other recent approaches~\cite{nercessian2021lightweight,colonel2022direct} have used the frequency sampling method during optimisation.
This approach allows a finite impulse response (FIR) approximation of an IIR filter's frequency response to be computed at arbitrary resolution, whilst retaining full differentiability.

\subsection{Our contribution}

In this work, we set out to address the shortcomings with existing neural network approaches using an end-to-end learning paradigm with differentiable signal processors.
Our proposed method can synthesise contact sounds from arbitrary 2D shapes and material parameters, without assuming any particular damping model, in a fraction of the time required for a numerical method.
Our network is coupled with a differentiable IIR filter bank which can be configured with varying arrangements of cascaded and parallel biquads.

\section{Method}
\label{sec:method}

Fig.~\ref{fig:pipeline} provides an overview of our method, illustrating how we sample synthetic data, prepare inputs, and train our model.


\subsection{Learnable modules}

To allow efficient synthesis of contact sounds at multiple object positions for a given shape, our system contains two separate neural networks which are jointly optimised.
The first is a convolutional neural network (CNN), using the EfficientNet-B0 architecture~\cite{tan2019efficientnet}, which generates an embedding from two-dimensional shapes.
The network takes as input a 2D occupancy grid describing the shape of an object in discretized space and outputs an embedding of 1000 dimensions.
For each distinct shape, only a single forward pass through this module is required.


The second learnable module is a multi-layer perceptron (MLP) which takes as input the shape embedding concatenated with material parameters and the co-ordinates of the vertex at which the shape will be excited.
The cached output of the CNN can be reused to evaluate different excitation positions and materials for the same shape, eschewing the need to recompute a shape embedding.
The material parameters $\boldsymbol\phi = (\rho, E, \nu, \alpha, \beta)$ consist of mass density $(\rho)$, Young's modulus ($E$), Poisson's ratio ($\nu$), and Rayleigh damping coefficients ($\alpha, \beta$).
Both excitation co-ordinates $\mathbf{x}$ and material parameters $\boldsymbol\phi$ are normalized in the range [0, 1].
The MLP's output is used to parameterise a differentiable IIR filter bank.

\subsection{Differentiable filter bank}

Each biquad in our differentiable filter bank is parameterized by a single pole $p_{l,m}$ and zero $q_{l,m}$, where $l$ and $m$ correspond to the filter's index in the parallel and cascade dimensions, respectively. We constrain the remaining pole and zero of each filter to the complex conjugates of these parameters.
Each biquad also has a gain parameter $k_{l,m}$ applied to its numerator.
In order to encourage filters with strong resonant peaks, we specify a fixed bias for pole and zero parameters, initialising them near the edge and centre, respectively, of the unit circle.

To ensure filter stability, we constrain the complex poles to be within the unit circle following the method described in \cite{nercessian2020neural}.
Specifically, we apply an activation function $h:\mathbb{C}\to\mathbb{C}$ defined as:

\begin{equation}
    \label{eqn:pole_activation}
    h(p)=\frac{\tanh|p|}{|p|}\cdot p
\end{equation}

\noindent We do not apply such a constraint to the zeros as we are unconcerned with preserving minimum phase.
The transfer function of each filter is thus given by:

\begin{equation}
    \label{eqn:biquad_transfer}
    H_{l,m}(z)=\frac{
    1 - 2\mathfrak{Re}(q_{l,m})z^{-1}+\left|q_{l,m}\right|^2z^{-2}
    }{
    1 - 2\mathfrak{Re}(h\left(p_{l,m}\right))z^{-1}+\left|h\left(p_{l,m}\right)\right|^2z^{-2}
    }
\end{equation}

While training the model, we do not evaluate the filter's exact recursive implementation, and instead opt, as in recent work on differentiable IIR modelling~\cite{nercessian2021lightweight, colonel2022direct}, to approximate the filter response by sampling it in the frequency domain.
That is, we evaluate the filter's transfer function at discrete points $e^{j2\pi \frac{k}{N}}$ on the unit circle, where $N$ is the signal length and $k\in\{0, \dots, N/2 + 1\}$.
Whilst this method incurs a loss in precision, it allows the filter computation to be performed in parallel on a GPU, whilst also sidestepping the aforementioned issues with training stability and exploding gradients.

Our filter bank consists of a flexible number of parallel filters, each consisting of a cascade of some number of second order (biquadratic) sections.
The cascade depth $M$ and number of parallel filters $L$ are thus tunable hyperparameters of our method, with the overall transfer function of the filter bank given by:

\begin{equation}
    \label{eqn:filterbank_transfer}
    H(z) =
    \sum_{l=1}^L
    \prod_{m=1}^M
    k_{l,m}
    H_{l,m}(z)
\end{equation}

\subsection{Loss}



We train our model to minimize the following loss function:

\begin{equation}\label{eqn:optim_objective}
\mathcal{L} =
\lambda \left\|
X_\text{mel} - H_\text{mel}
\right\|_2^2
+ \gamma \left\|
\log X_\text{mel} - \log H_\text{mel}
\right\|_2^2,
\end{equation}

\noindent where $\lambda$ and $\gamma$ are loss scaling hyperparameters, and $X_\text{mel}$ and $H_\text{mel}$ are the mel-scaled magnitude spectra of the discrete Fourier transform (DFT) and filter frequency response, respectively.
In our experiments we used $\lambda=1.0$, and $\gamma=0.1$.

\subsection{Dataset}

To train our model, we generated a synthetic dataset consisting of 500 convex shapes~\cite{valtr1995probability} along with their material parameters and audio synthesised using modal synthesis.

In particular, each convex shape was represented by a point set $P$ which was triangulated using Delaunay triangulation. Subsequently, each triangular surface mesh $\mathcal{M}=(V,E,F)$ with a set of $V$ vertex positions, $E$ edges and $F$ faces was rasterized into a $64\times 64$ occupancy grid.
Materials were homogeneous and isotropic, and their parameters were sampled uniformly in the intervals $\rho \in [500, 15000]$, $E \in [8\times 10^9, 5\times 10^{10}]$, $\nu \in [0.1, 0.4]$, $\alpha \in [1, 10]$ and $\beta \in [3\times 10^{-7}, 2\times 10^{-6}]$. For each shape, 500 random materials were sampled.

To assemble the stiffness and density matrices and solve the generalized eigenvalue problem corresponding to each shape-material pair, we used the scikit-fem~\cite{skfem2020} package.
We computed the first 32 modes for each object and projected a unit impulse for each node in the mesh.
Since each mesh has a distinct tessellation, the number of triangles per mesh varied between approximately 300 and 1000.
Each vertex thus represents a unique combination yielding a total of approximately 100 million samples.


\subsection{Training}

We trained our models using the Adam optimizer with an initial learning rate of $3 \times 10^{-5}$ and an exponential decay of $0.9$ every 300 steps.
Models were trained until validation loss did not improve for 20k steps.
All models were trained with a batch size of 64 on a single NVIDIA RTX A5000 GPU.
\section{Results}
\label{sec:experiments}

\begin{figure}[t]
    \centering
    \includegraphics[width=\columnwidth]{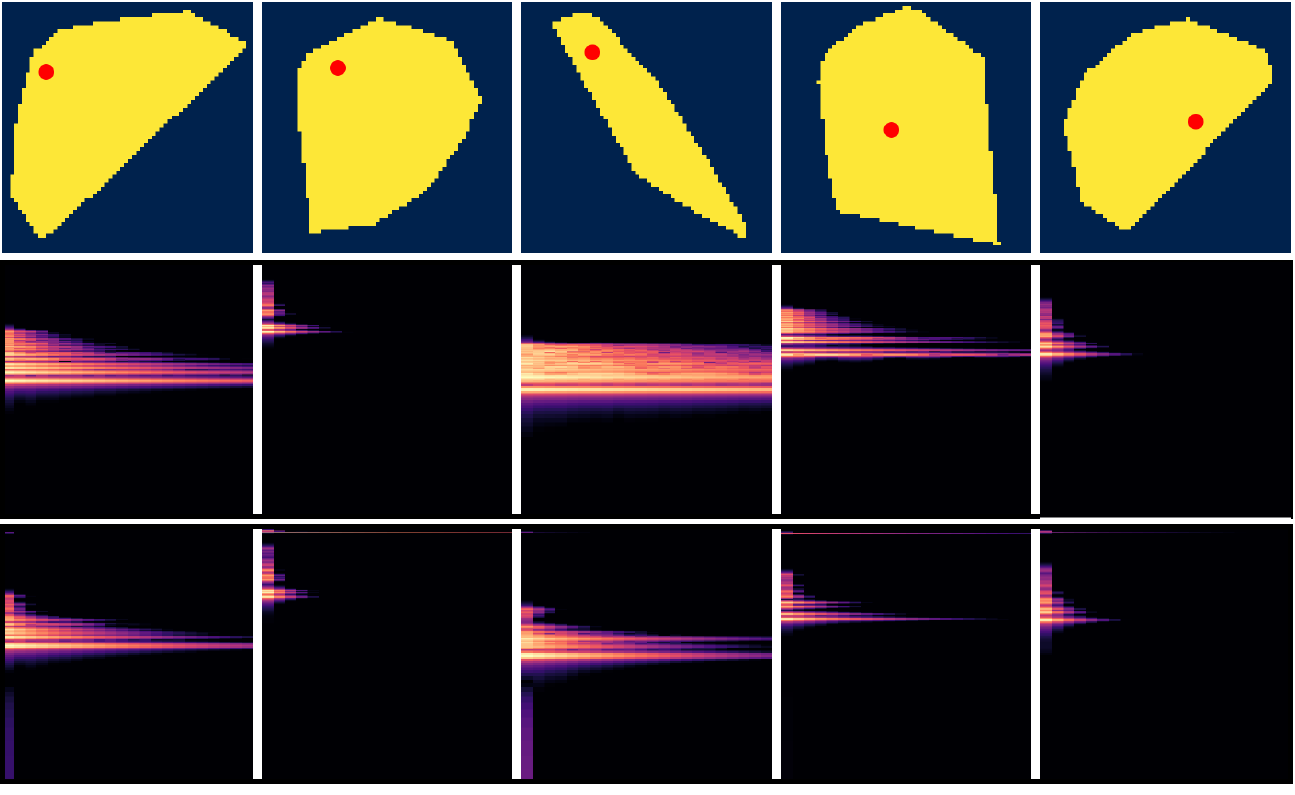}
    \caption{Inference results for test shapes, materials and positions. Top: Occupancy maps and indicated hit position in red. Middle: FEM results' spectrogram using traditional modal synthesis. Bottom: Results using our proposed method. Here the log-frequency log-magnitude spectrograms show that the generated results closely resemble the original approach.}
    \label{fig:results}
\end{figure}



In Fig.~\ref{fig:results} we illustrate our network's predictions for samples from a test dataset of 100 materials and shapes. 
We observe that our method closely matches the frequency distribution of the target signals, while generalising well across shapes.
We provide comprehensive examples of audio rendered with our system in the online supplement at \url{http://rodrigodzf.github.io/iir-modal/results.html}, including audio from the objects in Fig.~\ref{fig:results}, and interpolation of shape and material parameters.

Our model appears to produce shorter decay times than those in the target signal for modes of higher frequency.
Further, we note that the complex activation in Eqn~\ref{eqn:pole_activation} causes the gradient magnitude to decrease as a pole approaches the unit circle.
It is thus possible that for higher frequency modes with lower amplitudes, this diminishing gradient slowed learning of these resonances resulting in faster decays.

\subsection{Computation time}
To ascertain our method's computational performance, we compared its processing time to that of a conventional FEM solver (as implemented in \textit{scikit-fem}).
Table~\ref{tab:timing} lists the processing times of both methods for 96, 426, and 1792 vertices averaged across 100 meshes of different resolutions.

Our method produces filter bank parameters for a set of 96 arbitrary co-ordinates in a mesh in 16.75 milliseconds using a GPU.
Subsequently rendering the audio is possible at negligible computational cost using a recursive implementation of the filter bank.
In comparison, modal synthesis using the FEM solver takes approximately 83.66 milliseconds to calculate the mode gains and frequencies for a mesh with the same number of vertex positions.
The solutions are only available at the discrete coordinate positions of the vertices. Increasing the number of mesh vertices, and thus obtaining solutions at finer steps, increases the duration of processing.

To produce a mesh with 1792 vertices, the FEM solver takes slightly over 1.7 seconds, while our method requires only 0.08 seconds. Note that in these experiments we generate a shape embedding even though the occupancy grid does not necessarily change in the model input, and thus this can be considered a worst case scenario bound on the actual runtime of our model. If we cache the shape embedding, the total time is reduced to 0.01 seconds for this number of vertices.
Moreover, as our method does not need a re-discretization of the shape, we can query filter parameters at any continuous valued co-ordinate in the shape with the same constant processing time. Further, our method is not restricted to simultaneously querying multiple points in the mesh to render the sound for a single position. Our approach is therefore preferable in real-time and interactive synthesis for arbitrary shapes. 

\begin{table}[]
\centering
\begin{tabular}{crrr}
\toprule
 & \multicolumn{1}{c}{96 vertices}  & \multicolumn{1}{c}{426 vertices} & \multicolumn{1}{c}{1792 vertices} \\ \midrule
FEM             & 83.66ms & 387.12ms & 1728.25ms \\
Ours            & 16.75ms & 22.03ms  & 85.86ms \\
\bottomrule
\end{tabular}
\caption{Processing time in milliseconds for the FEM solver and our method. We average the processing time for 100 meshes of different resolutions. For the solver, we measure the time required to solve the system for each mesh. In our case, we add the inference time for a shape and material pair and the inference time for a fixed number of coordinate positions.}
\label{tab:timing}
\end{table}

\subsection{Filter-bank configuration}
To ascertain the optimal trade off between the cascade depth $M$ and number of parallel filters $L$, we compared three filter bank configurations such that the total number of poles was constant, fixed at $M\cdot L=64$.
We evaluated models trained using these configurations on the test dataset using both the frequency domain mean-absolute and mean-squared error.
In general, we observed that filter banks with a lower cascade depth and thus a lower filter order converged faster.

Table \ref{tab:filter_settings} lists the results for these experiments.
Convergence was very poor for the highest order filter bank, with dramatically higher errors than either of the other configurations.
The \nth{4} order filter bank achieved the best performance across both metrics, suggesting a hybrid cascade/parallel topology is optimal.

\begin{table}[]
\centering
\begin{tabular}{crr}
\toprule
Filters and order           & MAE             & MSE     \\ \midrule
16 filters \nth{8} order    & 4.324           & 20.854          \\
32 filters \nth{4} order    & \textbf{0.504}  & \textbf{0.574}  \\
64 filters \nth{2} order    & 0.568           & 0.697           \\
\bottomrule
\end{tabular}
\caption{Average log-spectrogram MAE and MSE for different filter configurations. We test our model using a different number of filters in parallel of different order.}
\label{tab:filter_settings}
\end{table}

\section{Conclusion}
\label{sec:conclusion}

This paper presented a novel approach for synthesising contact sounds for rigid surfaces of arbitrary shape and material composition.
Our method generates parameters for a resonant IIR filter bank from shape, material, and excitation position inputs.
The generated filter bank is suitable for real-time implementation, allowing realistic contact sounds to be produced interactively using a variety of different excitation signals.


In this work, we limited shapes to two dimensions, but we believe that extending the proposed method to 3D objects simply requires adapting the shape encoder to use 3D convolutional layers. 
Additionally, a future adaption of our model could apply a discretization agnostic network~\cite{sharp2022diffusionnet} to learn from different surface representations. The rest of the pipeline would not require any alteration. 


We further consider that, adapting a method previously applied to generative models \cite{han2020wav2shape, zhang2017generative}, our trained model could also be used to optimise a second network for the task of predicting materials and shapes from sound.

Finally, since the predicted filter bank can be excited with any signal, we intend to explore the possibilities of complex interaction on the object's surface, along with a qualitative evaluation.




\bibliographystyle{IEEEbib}
\bibliography{strings,refs}

\begin{thebibliography}{10}

\bibitem{o2002synthesizing}
James~F O'Brien, Chen Shen, and Christine~M Gatchalian,
\newblock ``Synthesizing sounds from rigid-body simulations,''
\newblock in {\em Proceedings of the 2002 ACM SIGGRAPH/Eurographics symposium
  on Computer animation}, 2002, pp. 175--181.

\bibitem{van2001foleyautomatic}
Kees Van Den~Doel, Paul~G Kry, and Dinesh~K Pai,
\newblock ``Foleyautomatic: physically-based sound effects for interactive
  simulation and animation,''
\newblock in {\em Proceedings of the 28th annual conference on Computer
  graphics and interactive techniques}, 2001, pp. 537--544.

\bibitem{rausch2017modal}
Dominik Rausch,
\newblock {\em Modal sound synthesis for interactive virtual environments},
\newblock Ph.D. thesis, Universit{\"a}tsbibliothek der RWTH Aachen, 2017.

\bibitem{bonneel2008fast}
Nicolas Bonneel, George Drettakis, Nicolas Tsingos, Isabelle Viaud-Delmon, and
  Doug James,
\newblock ``Fast modal sounds with scalable frequency-domain synthesis,''
\newblock in {\em ACM SIGGRAPH 2008 papers}, pp. 1--9. 2008.

\bibitem{chadwick2012faster}
Jeffrey~N Chadwick, Changxi Zheng, and Doug~L James,
\newblock ``Faster acceleration noise for multibody animations using
  precomputed soundbanks,''
\newblock in {\em Proceedings of the ACM SIGGRAPH/Eurographics Symposium on
  Computer Animation}, 2012, pp. 265--273.

\bibitem{traer2019perceptually}
James Traer, Maddie Cusimano, and Josh~H McDermott,
\newblock ``A perceptually inspired generative model of rigid-body contact
  sounds,''
\newblock in {\em The 22nd International Conference on Digital Audio Effects
  (DAFx-19)}, 2019.

\bibitem{jin2020deep}
Xutong Jin, Sheng Li, Tianshu Qu, Dinesh Manocha, and Guoping Wang,
\newblock ``Deep-modal: real-time impact sound synthesis for arbitrary
  shapes,''
\newblock in {\em Proceedings of the 28th ACM International Conference on
  Multimedia}, 2020, pp. 1171--1179.

\bibitem{jin2021deepeigen}
Xutong Jin, Sheng Li, Dinesh Manocha, and Guoping Wang,
\newblock ``Deepeigen: Learning-based modal sound synthesis with acoustic
  transfer maps,''
\newblock {\em arXiv preprint arXiv:2108.07425}, 2021.

\bibitem{jin2022neuralsound}
Xutong Jin, Sheng Li, Guoping Wang, and Dinesh Manocha,
\newblock ``Neuralsound: learning-based modal sound synthesis with acoustic
  transfer,''
\newblock {\em ACM Transactions on Graphics (TOG)}, vol. 41, no. 4, pp. 1--15,
  2022.

\bibitem{conan2014synthesis}
Simon Conan, Olivier Derrien, Mitsuko Aramaki, S{\o}lvi Ystad, and Richard
  Kronland-Martinet,
\newblock ``A synthesis model with intuitive control capabilities for rolling
  sounds,''
\newblock {\em IEEE/ACM transactions on audio, speech, and language
  processing}, vol. 22, no. 8, pp. 1260--1273, 2014.

\bibitem{kuznetsov2020differentiable}
Boris Kuznetsov, Julian~D Parker, and Fabi{\'a}n Esqueda,
\newblock ``Differentiable iir filters for machine learning applications,''
\newblock in {\em Proc. Int. Conf. Digital Audio Effects (eDAFx-20)}, 2020, pp.
  297--303.

\bibitem{pepe2020designing}
Giovanni Pepe, Leonardo Gabrielli, Stefano Squartini, and Luca Cattani,
\newblock ``Designing audio equalization filters by deep neural networks,''
\newblock {\em Applied Sciences}, vol. 10, no. 7, pp. 2483, 2020.

\bibitem{nercessian2021lightweight}
Shahan Nercessian, Andy Sarroff, and Kurt~James Werner,
\newblock ``Lightweight and interpretable neural modeling of an audio
  distortion effect using hyperconditioned differentiable biquads,''
\newblock in {\em ICASSP 2021-2021 IEEE International Conference on Acoustics,
  Speech and Signal Processing (ICASSP)}. IEEE, 2021, pp. 890--894.

\bibitem{colonel2022direct}
Joseph~T Colonel, Christian~J Steinmetz, Marcus Michelen, and Joshua~D Reiss,
\newblock ``Direct design of biquad filter cascades with deep learning by
  sampling random polynomials,''
\newblock in {\em ICASSP 2022-2022 IEEE International Conference on Acoustics,
  Speech and Signal Processing (ICASSP)}. IEEE, 2022, pp. 3104--3108.

\bibitem{tan2019efficientnet}
Mingxing Tan and Quoc Le,
\newblock ``Efficientnet: Rethinking model scaling for convolutional neural
  networks,''
\newblock in {\em International conference on machine learning}. PMLR, 2019,
  pp. 6105--6114.

\bibitem{nercessian2020neural}
Shahan Nercessian,
\newblock ``Neural parametric equalizer matching using differentiable
  biquads,''
\newblock in {\em Proc. Int. Conf. Digital Audio Effects (eDAFx-20)}, 2020, pp.
  265--272.

\bibitem{valtr1995probability}
P~Valtr,
\newblock ``Probability that n random points are in convex position,''
\newblock {\em Discrete \& Computational Geometry}, vol. 13, no. 3-4, pp.
  637--643, 1995.

\bibitem{skfem2020}
Tom Gustafsson and G.~D. McBain,
\newblock ``scikit-fem: A {P}ython package for finite element assembly,''
\newblock {\em Journal of Open Source Software}, vol. 5, no. 52, pp. 2369,
  2020.

\bibitem{sharp2022diffusionnet}
Nicholas Sharp, Souhaib Attaiki, Keenan Crane, and Maks Ovsjanikov,
\newblock ``Diffusionnet: Discretization agnostic learning on surfaces,''
\newblock {\em ACM Transactions on Graphics (TOG)}, vol. 41, no. 3, pp. 1--16,
  2022.

\bibitem{han2020wav2shape}
Han Han and Vincent Lostanlen,
\newblock ``wav2shape: Hearing the shape of a drum machine,''
\newblock {\em arXiv preprint arXiv:2007.10299}, 2020.

\bibitem{zhang2017generative}
Zhoutong Zhang, Jiajun Wu, Qiujia Li, Zhengjia Huang, James Traer, Josh~H
  McDermott, Joshua~B Tenenbaum, and William~T Freeman,
\newblock ``Generative modeling of audible shapes for object perception,''
\newblock in {\em Proceedings of the IEEE International Conference on Computer
  Vision}, 2017, pp. 1251--1260.

\end{thebibliography}

\end{document}